# Illness-death model with renewal

Ralph Brinks

Medical Biometry and Epidemiology, Witten/Herdecke University, Faculty of Health/School of Medicine, 58448 Witten, Germany

Institute for Biometry and Epidemiology, German Diabetes Center, 40225 Düsseldorf, Germany

## Abstract

The illness-death model for chronic conditions is combined with a renewal equation for the number of newborns taking into account possibly different fertility rates in the healthy and diseased parts of the population. The resulting boundary value problem consists of a system of partial differential equations with an integral boundary condition. As an application, the boundary value problem is applied to an example about type 2 diabetes.

## Keywords

## Introduction

Chronic diseases cause the highest public health burden worldwide with an increasing trend over time: an estimated percentage of 61% of global deaths (equivalent to 31 million) in 2000 and 74% (equivalent to 41 million) in 2019 were attributable to chronic diseases [WHO23]. Despite the huge burden, attempts to quantitatively model the progression of chronic conditions and impact of risk factors are scarce compared to the manifold approaches and the long tradition of modeling infectious diseases [Chu10].

As a contribution to filling this gap, we have proposed a system of partial differential equations (PDEs) to describe the number of people transiting through the illness-death model (IDM) for chronic conditions [Bri14, Bri15]. The IDM for chronic conditions is a compartment model that divides the population under consideration into the states *Non-diseased* and *Diseased*. *Diseased* in this context means suffering from a specific chronic condition. Later in the text, the chronic condition will be type 2 diabetes. *Non-diseased* and *Diseased* in this case means without and with type 2 diabetes, respectively. To account for mortality from the states *Non-diseased* and *Diseased* an additional state *Dead* is also added to the IDM. The IDM for chronic diseases is shown in Figure 1, where possible transitions between the states are indicated as arrows.

The numbers of people in the states *Non-diseased* and *Diseased* henceforth are called $S$ and $C$, respectively. Both numbers depend on the calendar time $t$ and the age $a$, which means that they are functions of $t$ and $a$: $S = S(t, a)$ and $C = C(t, a)$. The numbers $S$ and $C$ can be related to the transition rates in the IDM. Transition rates (synonymously: transition densities or transition intensities) are the incidence rate $i = i(t, a)$ and the mortality rates $m_0 = m_0(t, a)$ and $m_1 = m_1(t, a)$. The indices 0 and 1 of the mortality rates denote if someone dies without ($m_0$) or with ($m_1$) the chronic condition.

In [Bri15] we have shown that in case there is no migration into the states *Non-diseased* and *Diseased*, the numbers $S$ and $C$ are described by the following PDEs:

$$\partial S = -(i + m_0)\, S \qquad (1a)$$
$$\partial C = i\, S - m_1\, C \qquad (1b)$$

where for $X \in \{S, C\}$ the symbol $\partial X$ denotes the directional derivative in the (1,1)-direction

$$\partial X(t, a) = \lim_{h \to 0} \frac{X(t+h,\, a+h) - X(t, a)}{h} \qquad (2)$$

In case, $S$ and $C$ are sufficiently smooth, the directional derivative in Equation (2) is the sum of the partial derivatives with respect to time $t$ and age $a$, i.e. $\partial X = \partial X/\partial t + \partial X/\partial a$.

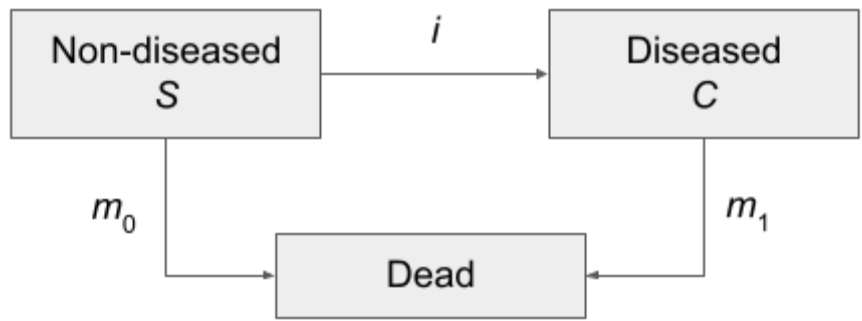

Figure 1: Illness-death model for chronic conditions. The population is split into the states *Non-diseased* and *Diseased*. To account for mortality, a third state *Dead* is added. Possible transitions between the states are indicated as arrows; associated transition rates are the incidence rate $i$ and the mortality rates $m_0$ and $m_1$.

To treat and uniquely solve the PDE system given by Eqs (1a) and (1b), initial or boundary conditions are necessary. Frequently, it is assumed that the subjects of the considered population contract the chronic condition only after birth, which implies $C(t, 0) = 0$ for all $t$. In this case, an important question is what the initial condition for the (healthy) newborns $S(t, 0)$ is. Sometimes, we can make assumptions about the numbers of newborns, for example, in demographic projections where future fertility in the population is dealt with in different scenarios [Pos19]. In fact, for a variety of projections about future case numbers of chronic conditions, we used this approach: The number of newborns stemmed from the population projections of the national statistical offices [Toe19, Toe23, Voe23, Wan23]. In this article, we will describe a new approach including the reproductive behavior of the population itself. For this, we will keep the assumptions of 1) no migration and 2) that newborns are disease-free, and will formulate a renewal equation for the birth process.

# Methods

We will extend the Sharpe-Lotka-McKendrick's PDE for the IDM and model the birth process by a renewal law similar to Eq. (4.2) of [Ani11] in terms of *S* and *C*. For this, we will use two fertility rates $\beta$ and we allow that the fertility rate of people in the *Non-Diseased* and *Diseased* states are different. The associated rates are denoted by $\beta_0$ and $\beta_1$. Again the sub-indices indicate whether the fertility rates refer to non-diseased (sub-index 0) and diseased (1) subjects. For example, in type 2 diabetes the fertility rate of people with diabetes is estimated to be lowered by (about) 25% compared to people without diabetes: $\beta_1(t, a) = 0.75\,\beta_0(t, a)$ [Mat21]. In our diabetes example, the fertility rates $\beta_0$ and $\beta_1$ are non-zero for ages *a* between 15 and 45 years; outside this interval, the birth rates are assumed to be 0.

Apart from the renewal law for $S(t, 0)$ and $C(t, 0)$, $t > 0$, we will give boundary conditions for $S(0, a)$ and $C(0, a)$ for $a \geq 0$. Using the method of characteristics [Pol02], the resulting boundary value problem can be treated by reducing it to a system of ordinary differential equations (ODEs), which is solved along the characteristic lines in the (1, 1)-direction. The situation is shown in Figure 2. We are interested in the solution of the PDE system given by Eqs. (1a) and (1b) in the gray rectangular region of Figure 2, which is a subset of the Lexis plane [Kei90]. The two types of initial conditions are indicated by red lines: along the solid red line *S* and *C* take prescribed values, and along the dashed red line the initial condition is defined by the renewal equation using the birth rates $\beta_0$ and $\beta_1$. Given the initial conditions, the PDEs as in Eqs. (1a) and (1b) are solved as ODEs along the characteristic lines in the (1, 1)-direction; four of these are shown in Figure 2 as solid black lines.

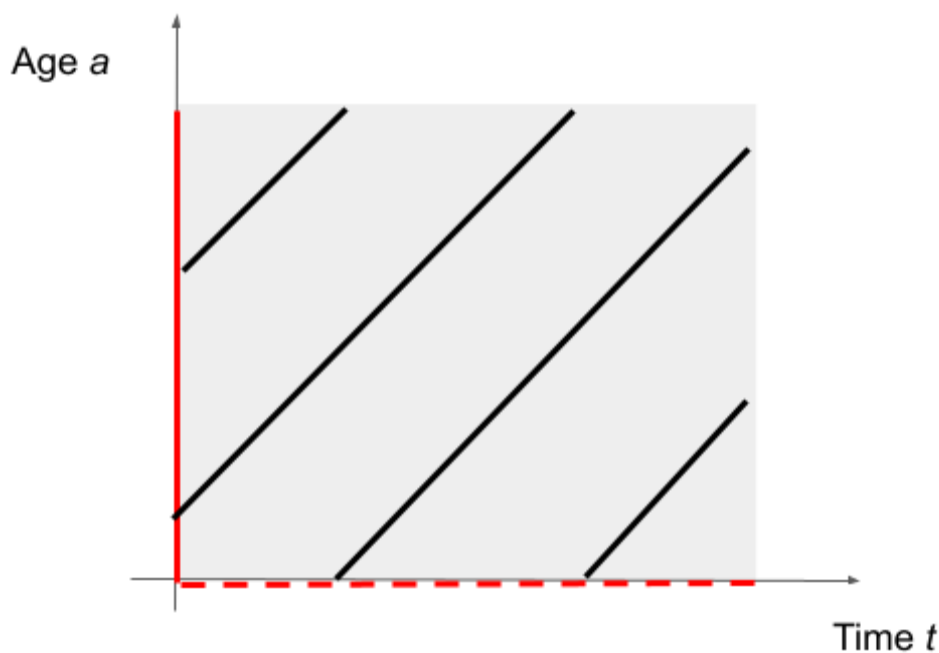

Figure 2: Definition of the boundary conditions (red) and characteristic lines (black) in the Lexis plane. Two types of initial conditions are used: along the solid red line *S* and *C* take prescribed values; along the dashed red line the initial condition is defined by the renewal equation. The PDEs as in Eqs. (1a) and (1b) are solved along the characteristic lines; four of these are shown in black.

The initial condition along the vertical line in Figure 2, i.e., *S*(0, *a*) and *C*(0, *a*), is chosen by a prescribed prevalence *p*(*a*), 0 ≤ *p*(*a*) ≤ 1, and setting *S*(0, *a*) = 1000 x (1 − *p*(*a*)) and *C*(0, *a*) = 1000 x *p*(*a*). The age course of the prevalence is motivated by type 2 diabetes in Germany [Tam16]. Then, we solve the system (1) in the rectangle (*t*, *a*) ∈ [0, 50] x [0, 100] using the Runge-Kutta method of fourth order (RK4) [Dah74] for the resulting system of ODEs along the characteristic lines. Incidence and mortality rates again are motivated by type 2 diabetes in Germany.

To show the effect of the decreased birth rate $\beta_1$ we compare the scenario $\beta_1$ = 0.75 $\beta_0$ with the reference scenario of equal birth rates $\beta_1$ = $\beta_0$; the respective scenarios are denoted by B1 and B0. The prescribed prevalence *p*(*a*) is the same in both scenarios B0 and B1.

All calculations are done with the free statistical software R (The R Foundation for Statistical Computing). The source code for the calculations is available in the public repository Zenodo under DOI 10.5281/zenodo.10137926.

## Results

To formulate an initial condition for the system given by Eqs. (1a) and (1b) with all newborns being disease free, i.e. *C*(*t*, 0) = 0 for all *t*, we have

$$S(t, 0) = \int_0^\infty \left[\beta_0(t, a) S(t, a) + \beta_1(t, a) C(t, a)\right] da \quad (3)$$

where $\beta_0$(*t*, *a*) and $\beta_1$(*t*, *a*) are the reproduction rates of the non-diseased and diseased subjects aged *a* at time *t*, respectively. Equation (3) is a renewal equation following the same construction as Eq. (4.2) of [Ani11]. In our case, we have two different parts of the population shown in Figure 1, one part being diseased, the other being undiseased; the associated numbers of these two parts are described by *S* and *C* and the associated birth rates are $\beta_0$ and $\beta_1$.

Figure 3 shows the numbers of newborns *S*(*t*, 0) in the two scenarios B0 and B1 over time. In scenario B1 (reduced birth rate in diseased people) the number of newborns is approximately 2% lower over the years compared to scenario B0 (same birth rates for healthy and diseased people).

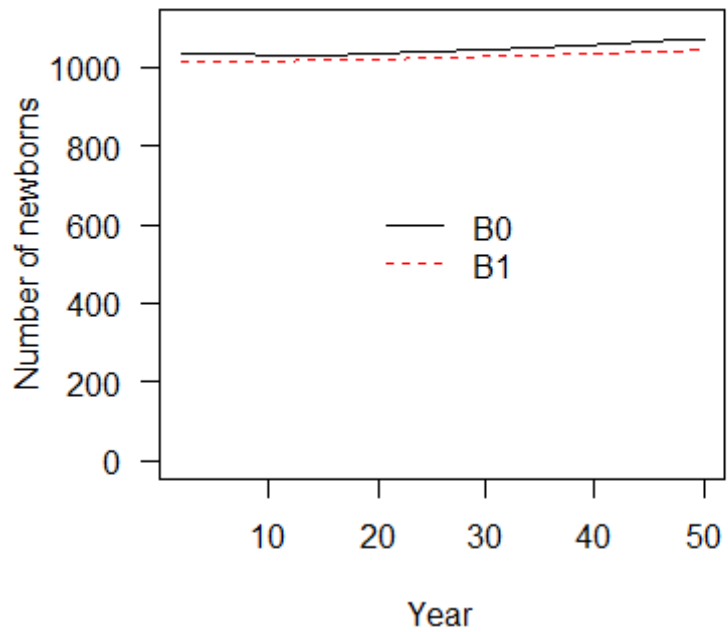

Figure 3: Number of newborns $S(t, 0)$ over time in the two scenarios B0 (without reduced fertility, black line) and B1 (with reduced fertility, dashed red). The decreased birth rate in the group of diseased people leads to an reduction of about 2% in newborns over the years.

Figure 4 presents the age-specific number of people $C(t, a)$ for $t = 10$, 30 and 50 in the range $a \in [0, 100]$. The left and right panels of Figure 4 show the reference case B0 and the scenario B1, respectively. The graphs for the two scenarios are visually indistinguishable.

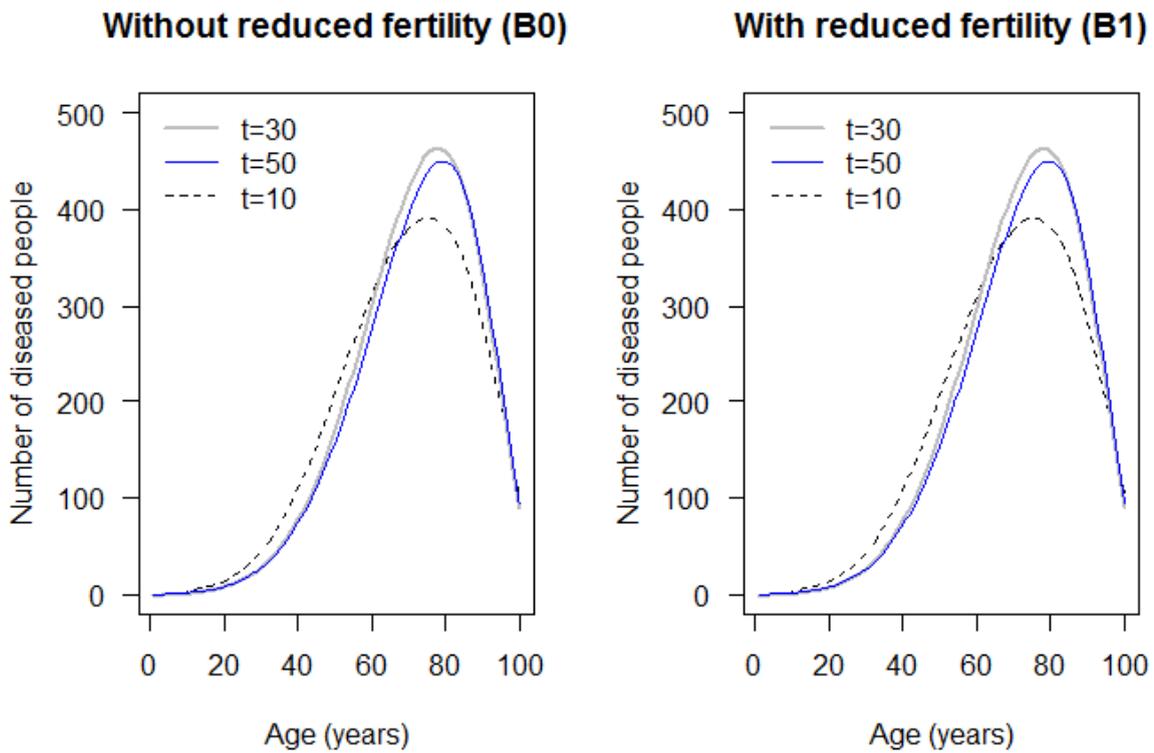

Figure 4: Numbers of diseased people *C* in the years *t* = 10 (dashed black), 30 (solid gray) and 50 (solid blue) over age in the two different fertility scenarios B0 (left panel) and B1 (right panel). Visually, there is no difference between the curves in the two scenarios.

For better comparison, for each year the differences in the numbers S and C between the scenarios B0 and B1 are calculated and presented in Figure 5. The difference is calculated as the total number of healthy and diseased people summed over all ages in scenario B0 minus the total number in scenario B1. The difference in the number of healthy people (shown in the left panel of Figure 5) is a magnitude higher than the number of ill people (right panel).

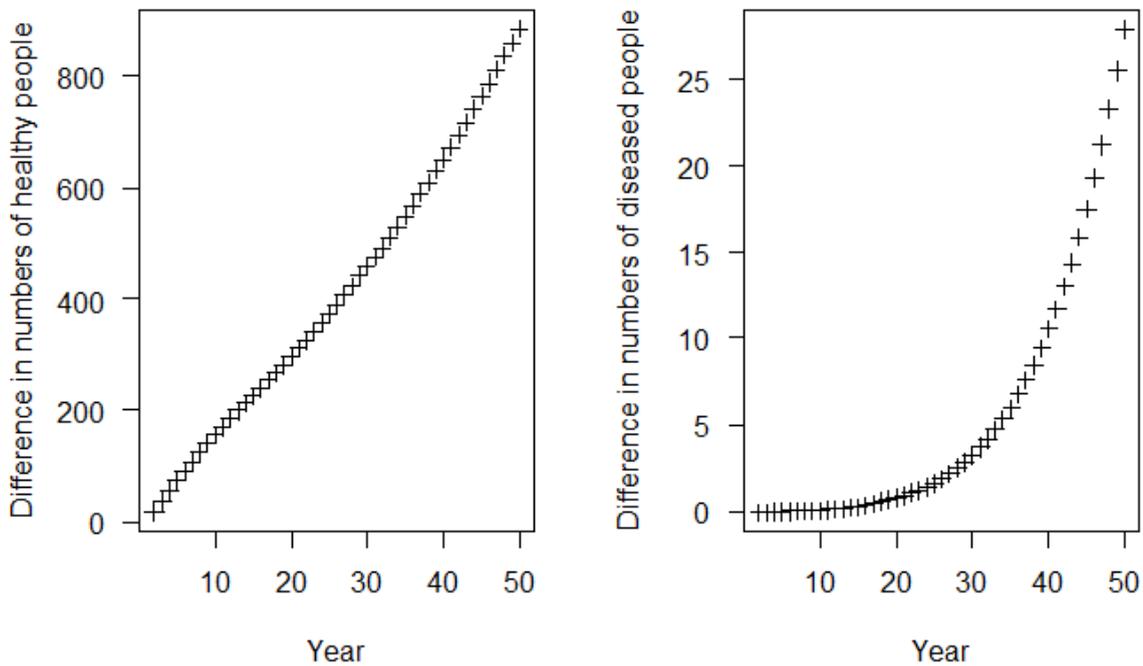

Figure 5: Differences in the total numbers of healthy (left panel) and diseased people (right) versus calendar time *t*. The difference is calculated for each year *t* as the total number (summed over all ages) in a year for scenario B0 minus the total number for scenario B1.

## Discussion

For the first time, we could estimate the amount of missing newborns due to reduced fertility in people with chronic conditions. This is possible by combining the equations (Eqs. (1a) and (1b)) describing the transitions in the illness-death model with a renewal equation (Eq. (2)), which can take into account the possibly reduced fertility rate in people with illness.

As an example, we have chosen type 2 diabetes in Germany, because a considerable part of the population is affected by this chronic disease. We must note that the part of the German population with type 2 diabetes and the population with interest in reproduction have little overlap in age. This can be different for other diseases, like for example type 1 diabetes or can be different in other populations. For instance in the US, type 2 diabetes is much more prevalent in the younger population [Toe23] such that the effect of reduced fertility in people with the chronic condition can have a higher impact than in the example for Germany presented in this article.

For simplicity, no interaction between the two sexes necessary for reproduction has been taken into account. Restricting reproductive age to the span 15 to 45 years of age widely reflects reality for the female population. In Germany, ages of becoming fathers occasionally lie above age 45. If mating and the possibly different ages of reproduction between women and men needs to be considered, two-sex models as used in the theory of sexually transmitted diseases (STDs) are necessary. For an introductory text about two-sex models, we refer to chapter 8 in [Vyn10].

# Declarations

The author declares that he has no competing interest regarding any aspect of this work. The author is employed at University Witten/Herdecke and at the German Diabetes Center (Düsseldorf). Apart from the salary by these two institutions, the author of this work did not receive any funding. No patients were enrolled for this work; the data are publicly available and properly cited. The source code for the calculations is available in the public repository Zenodo under DOI 10.5281/zenodo.10137926.